\documentstyle[12pt,epsfig]{article}
\input epsf
\begin{document}
\title{Phys. Rev. D {\bf 63}, 014017 (2001).\\[0.8cm]
Effective Lagrangian induced by the anomalous Wess-Zumino action and
the exotic resonance state with $I^G(J^{PC})=1^-(1^{-+})$ in the $\rho\pi$, $
\eta\pi$, $\eta'\pi$, and $K^*\bar K+\bar K^*K$ channels\\[1cm]}
\author{N.N. Achasov \thanks{Email address: achasov@math.nsc.ru}\,\, 
and G.N. Shestakov \thanks{Email address: shestako@math.nsc.ru}\\[0.8cm]
{\it Laboratory of Theoretical Physics,}\\
{\it S.L. Sobolev Institute for Mathematics,}\\ 
{\it 630090, Novosibirsk 90,Russia}\\[1cm]}
\date{} 
\maketitle\begin{abstract} A simple model for the exotic
waves with $I^G(J^{PC})=1^-(1^{-+})$ in the reactions $VP\to VP$,
$VP\to PP$, and $PP\to PP$ is constructed beyond the scope of the
quark-gluon approach. The model satisfies unitarity and
analyticity and uses as a ``priming"\ the ``anomalous"\ nondiagonal
$VPPP$ interaction which couples together the four channels
$\rho\pi$, $\eta\pi$, $\eta'\pi$, and $K^*\bar K+\bar K^*K$. The
possibility of the resonancelike behavior of the
$I^G(J^{PC})=1^-(1^{-+})$ amplitudes belonging to the $\{10\}-
\{\bar{10}\}$ and $\{8\}$ representations of $SU(3)$ as well as
their mixing is demonstrated explicitly in the $1.3-1.6$ GeV mass
range which, according to the current experimental evidence, is
really rich in exotics.
\end{abstract} \vspace*{0.8cm}

\,\ \ \ PACS number(s): 12.39.Fe, 13.25.Jx, 13.75.Lb
\newpage
\begin{center}{\bf I. INTRODUCTION}\end{center}

Phantoms of manifestly exotic states with $I^G(J^{PC})=1^-(1^{-+})$ have more 
and more agitated the experimental and theoretical communities [1-20]. They 
were discovered in the $1.3-1.6$ GeV mass range in the $\eta\pi$, $\eta'\pi$, 
$\rho\pi$, $b_1\pi$, and $f_1\pi$ systems produced in $\pi^-p$ collisions at 
high energies and in $N\bar N$ annihilation at rest [1-12].

Theoretical considerations concerning the mass spectra and decay properties of
exotic hadrons have been based, in the main, on the MIT-bag model, constituent
gluon model, flux-tube model, QCD sum rules, lattice calculations, and various
selection rules. The more recent discussions of these constituent quark-gluon
models and selection rules in conformity with the observed $J^{PC}=1^{-+}$
phenomena can be found in Refs. [1-20], together with extensive analyses of the
current experimental data and comprehensive references.
A resonance character of the observed exotic signals and also the more popular
assumption about their hybrid $(q\bar qg)$ nature are the subject of
much attention and require further careful investigations [1,6,7,12,14-20].

Let us note that evidence for the possible existence of an exotic $J^{PC}=1
^{-+}$ state coupled to the $\eta\pi$ and $\rho\pi$ channels and belonging to
the icosuplet representation of $SU(3)$ was obtained for the first time by
using the bootstrap technique of Schechter and Okubo more than 35 years ago
[21] (see also Ref. [22]).

Current algebra and effective chiral Lagragians are also important sources
of theoretical information on exotic partial waves. It is sufficient to
remember the prediction obtained within the framework of these approaches for
the $\pi\pi\,$ $s$-wave scattering length with isospin $I=2$ [23,24].
Constructing with the help of the effective chiral Lagragians the series
expansions of the scattering amplitudes in powers of external momenta, one can
reveal explicitly exotic contributions already among the lower order terms of
these series. Can
at least some of these contributions found at low energies turn out to be the
manifestations (``the tails") of high-lying exotic resonances? It is well
known that, for example, for the $\pi\pi$ scattering channels
involving the $\sigma$ and $\rho$ resonances, one can self-consistently (in the
sense of agreement with experiment) sew together the resonancelike and
low-energy behaviors of the scattering amplitudes by using the successfully
selected unitarization scheme for the original chiral contributions, together
with general analyticity requirements [25-34]. In other words, for these
channels, there exist a good many of the model constructions which show that
the low-energy contributions calculated within the effective chiral Lagragians
framework may in principle transform with increasing energy into resonances
with the experimentally established parameters. In the present work we continue
in this way and construct a model satisfying unitarity and analyticity for an
exotic wave with $J^{PC}=1^{-+}$ in the reaction $\rho\pi\to\eta\pi$ and in the
related reactions $\rho\pi\to\eta'\pi$, $\rho\pi\to\rho\pi$, $\eta\pi\to\eta\pi
$, $\eta\pi\to(K^*\bar K+\bar K^*K)$, and so on, using as a ``priming"\ the 
tree exotic amplitudes generated by a simplest ``anomalous"\ effective 
interaction of the vector ($V$) and pseudoscalar ($P$) mesons. The interaction 
is induced by the anomalous Wess-Zumino chiral Lagrangian [35] and is 
proportional to $\epsilon_{\mu\nu\tau\kappa}$.

At the tree level, the standard nonlinear chiral Lagrangian
describing the low-energy dynamics of the pseudoscalar mesons
belonging to the $SU(3)$ octet generates the $PP\to PP$ scattering
amplitudes possessing only the usual quantum numbers
$J^{PC}=0^{++}$ and $1^{--}$ in the $s$ channel [24]. However,
already in the next order of chiral perturbation theory, the
$J^{PC}=1 ^{-+}$ exotic contributions arise in these amplitudes at
the expense of the finite parts of the one-loop diagrams. In so
doing they turn out to be different from zero only owing to
$SU(3)$ symmetry breaking for pseudoscalar masses. The resonances,
with which such contributions might be associated, have to possess
rather odd properties. All their coupling constants to the octet
of pseudoscalar mesons must vanish in the $SU(3)$ symmetry limit.
Therefore, it seems more reasonable to assume that if the exotic
resonances with $J^{PC}=1^{-+}$ exist, then they are of another
origin. In such a case, the resources for their possible
generation, which still remain within the effective chiral
Lagrangian framework, seem to involve the ``anomalous"\
interactions of the vector and pseudoscalar mesons [36-43]. Some
indirect evidence in favor of this assumption has been given by
the analysis of the $PP\to PP$ scattering amplitudes carried out
in the framework of the linear $SU(3)\times SU(3)$ $\sigma$ model
involving only scalars and pseudoscalars [28]. There operate the
repulsive forces in the $J^{PC}=1^{-+}$ channels in this model,
and any resonance states do not arise.

In Sec. II, the general properties of the $VP\to PP$ reaction amplitudes are
briefly discussed within the framework of the unitary symmetry assumption. In
Sec. III, a simple model for the $I=1$ $p$-wave (exotic) reaction amplitudes
$VP\to PP$, $VP\to VP$, and $PP\to PP$ is constructed. The model takes into
account as a ``priming"\ the nondiagonal $VPPP$ interaction which couples
together the four channels $\rho\pi$, $\eta\pi$, $\eta'\pi$, and $K^*\bar K+
\bar K^*K$. It is essentially the summing up of all the $s$-channel loop
diagrams with the $VP$ and $PP$ intermediate states. In Sec. IV, the
possibility of the resonance-like
behavior of the $I^G(J^{PC})=1^-(1^{-+})$ amplitudes belonging to the $\{10\}-
\{\bar{10}\}$ and $\{8\}$ representations of $SU(3)$ as well as their mixing is
demonstrated explicitly in the $1.3-1.6$ GeV mass range. In quark-gluon
language, the $\{10\}-\{\bar{10}\}$ representation of $SU(3)$ first occurs in
the $qq\bar q\bar q$ sector, whereas the states with $J^{PC}=1^{-+}$ belonging
to the octet representation of $SU(3)$ may in principle correspond to both $qq
\bar q\bar q$ and $q\bar qg$ configurations.

\begin{center}{\bf II. GENERAL PROPERTIES OF THE \boldmath{$VP\to PP$}
AMPLITUDE}\end{center}

The general Lorentz and $SU(3)$ structure for the amplitude of the reaction $
V_a(k)+P_b(q_1)\to P_c(q_2)+P_d(q_3)$, where $V_a$ and $P_a$ are the members of
the vector and pseudoscalar octets taken in the Cartesian basis ($a=1,...,8$),
\footnote{$V_a=(\rho_1,\rho_2,\rho_3,K^*_4,K^*_5,K^*_6,K^*_7,\omega_8)$ and
$P_a=(\pi_1,\pi_2,\pi_3,K_4,K_5,K_6,K_7,\eta_8)$.} and $k,\,q_1,\,q_2$, and $q
_3$ are the four-momenta of the particles in the reaction, has the form
\begin{eqnarray}M^{(\lambda)}_{ab;cd}=-i\epsilon_{\mu\nu\tau\kappa}e_{(\lambda)
}^\mu q_1^\nu q_2^\tau q_3^\kappa\,[f_{abm}d_{mcd}A(s,t,u)+d_{abm}f_{mcd}B(s,t,
u)+(u_{ab})_{cd}C(s,t,u)].\end{eqnarray} Here $f_{abc}$ and $d_{abc}$ are the
standard structure constants of $SU(3)$ [44], $(u_{ab})_{cd}=f_{cam}d_{mbd}-d
_{cam}f_{mbd}$ [45], $e_{(\lambda)}^\mu$ is a $\mu$ component of the $V$ meson
polarization vector with helicity $\lambda$,\  $s=(k+q_1)^2$,\  $t=(k-q_2)^2$, 
and $u=(k-q_3)^2$. From Bose symmetry it follows that the invariant amplitude
$A(s,t,u)$ is antisymmetric under the interchange of the $t$ and $u$ variables,
whereas the invariant amplitudes $B(s,t,u)$ and $C(s,t,u)$ are symmetric. Note
that Eq. (1) can be obtained in the usual way [45-48] by applying $SU(3)$
symmetry, together with $P$ and $C$ invariance.

The first and second terms in Eq. (1) correspond to the octet transition
amplitudes $\{8_a\}\to\{8_s\}$ and $\{8_s\}\to\{8_a\}$ which we shall designate
for short by $A_{as}$ and $A_{sa}$, respectively; as usual, $\{8_s\}$ and $\{8
_a\}$ mean the symmetric and antisymmetric octet representations of $SU(3)$
which occur in the direct production of $\{8\}\times\{8\}$. The third term in
Eq. (1) describes transitions via the mutually conjugate representations $\{10
\}$ and $\{\bar{10}\}$ with the amplitudes $A_{10}$ and $A_{\bar{10}}$
appearing in the combination $A_{10}-A_{\bar{10}}$. In other words, it
describes the transitions from the initial $VP$ icosuplet $\{10\}-\{\bar{10}\}$
to the final $PP$ icosuplet $\{10\}+\{\bar{10}\}$. The transition amplitudes
between the self-conjugate representations $A_{as}$ and $A_{sa}$ can be
expanded into the partial waves with $J^{PC}=2^{++},4^{++},...$ and $J^{PC
}=1^{--},3^{--},...$, respectively. Hence they do not contain any explicitly
exotic contributions. As for the $\eta_8\pi$ final state, it dos not occur in
the $\{8_a\}$ but can belong to the representations $\{8_s\}$, $\{10\}$, and $
\{\bar{10}\}$ [21,22,49]. The $SU(3)$ exotic meson amplitudes $A_{10}$ and $A_{
\bar{10}}$ can be expanded into partial waves with $J^P=1^-,3^-,...$ . The
isotriplet amplitudes of $A_{10}-A_{\bar{10}}$ correspond to two sets of the
reactions with
opposite $G$ parity in the $s$ channel: (a) $\rho\eta_8\to\pi\pi$, $\rho\eta_8
\to K\bar K$, $\omega_8\pi\to\pi\pi$, $\omega_8\pi\to K\bar K$ and (b) $\rho\pi
\to\eta_8\pi$, $K^*\bar K\to\eta_8\pi$, $\bar K^*K\to\eta_8\pi$. The partial
amplitudes of the reactions belonging to set (a) possess the nonexotic quantum
numbers $I^G(J^{PC})=1^+(1^{--},3^{--},...)$ and, therefore, only hidden $SU(3)
$ exotics. The reactions belonging to set (b) are purely exotic because they
contain the partial waves with $I^G(J^{PC})=1^-(1^{-+},3^{-+},...)$. In
particular, it is these reactions that will be the subject of our attention in
the following.

Let us now write down the amplitude for the reaction $V_a(k)+P_b(q_1)\to P_c(q
_2)+P_d(q_3)$ involving the $V_0$ and $P_0$ $SU(3)$ singlets: \begin{eqnarray}
N^{(\lambda)}_{ab;cd}
=-i\epsilon_{\mu\nu\tau\kappa}e_{(\lambda)}^\mu q_1^\nu q_2^\tau q_3^\kappa\,[
\delta_{a0}f_{bcd}D(s,t,u)+\delta_{b0}f_{acd}E(s,t,u)+\delta_{c0}f_{abd}F(s,t,
u)]\,,\end{eqnarray} where $a,b,c,d$ are the flavor indices running now over $0
,1,...,8$, $f_{ab0}=0$, $V_0=\omega_0$, and $P_0=\eta_0$. By the isoscalar
particles with the definite masses we shall mean the pseudoscalar mesons $\eta=
\eta_8\,\cos\theta_P-\eta_0\,\sin\theta_P$ and $\eta'=\eta_8\,\sin\theta_P+\eta
_0\,\cos\theta_P$ with the mixing angle $\theta_P\approx-20^\circ$ [50,51] and
the vector mesons $\omega=\sqrt{1/3}\,\omega_8+\sqrt{2/3}\,\omega_0$ and $\phi=
\sqrt{2/3}\,\omega_8-\sqrt{1/3}\,\omega_0$ with ``ideal mixing". Equation (2)
describes the transitions via the $SU(3)$ octet intermediate states. The first
two terms in Eq. (2) do not contribute to $\eta\pi$ and $\eta'\pi$ production
because they correspond to the transitions into the final states belonging to
the $\{8_a\}$ representation which does not contain the $\eta_8\pi$ system. The
third term in Eq. (2) describes $\eta\pi$ and $\eta'\pi$ production via the $SU
(3)$ singlet components of the $\eta$ and $\eta'$. Under $t
\leftrightarrow u$ interchange, the invariant amplitudes $D(s,t,u)$ and $E(s,t,
u)$ are symmetric, while the invariant amplitude $F(s,t,u)$ does not possess
a definite symmetry. Thus, the first two terms in Eq. (2) can be expanded into
partial waves with $J^{PC}=1^{--},3^{--},...$ and the third term into 
partial waves with $J^{PC}=1^{-+},2^{++},3^{-+},4^{++},...$, of which the odd
waves $1^{-+},3^{-+},...$ are exotic. In principle, Eqs. (1) and (2) permit
the $VP\to PP$ reaction channels involving the $\omega$, $\phi$, $\eta$, and $
\eta'$ mesons to be considered in the most general form.

\begin{center}{\bf III. MODEL FOR THE  $I^G(J^{PC})=1^-(1^{-+})$ 
WAVES IN THE REACTIONS \boldmath{$VP\to PP,\ \,PP\to PP$}, AND \boldmath{$VP\to
VP$}}\end{center}

Consider the $SU(3)$ symmetric effective Lagrangian for the pointlike $VPPP$
interaction which also possesses additional nonet symmetry with respect to the
vector mesons, \begin{eqnarray}L(VPPP)=ih\,\epsilon_{\mu\nu\tau\kappa}
\mbox{Tr}(\hat V^\mu\partial^\nu\hat P\partial^\tau\hat P\partial^\kappa\hat P)
+i\sqrt{1/3}\,h'\,\epsilon_{\mu\nu\tau\kappa}\mbox{Tr}(\hat V^\mu\partial^\nu
\hat P\partial^\tau\hat P)\partial^\kappa\eta_0\,,\end{eqnarray} where $\hat
P=\sum_{a=1}^{8}\lambda_aP_a/\sqrt{2}\,$, $\hat V^\mu=\sum_{a=0}^{8}\lambda_a
V^\mu_a/\sqrt{2}\,$, and $\lambda_a$ are the Gell-Mann matrices [44]. The tree
amplitudes for the reactions $VP\to PP$ generated by the Lagrangian (3) are
given by Eqs. (1) and (2) with the following sets of the invariant amplitudes
(here we omit their arguments for short): $(A,B,C,D)=h(0,2,1,\sqrt{6})$
and $(E,F)=h'(\sqrt{2/3},-\sqrt{2/3})$. The presence of the amplitudes $C$ and
$F$ such as above implies (see the discussion in Sec. II) that the Lagrangian
(3) generates tree exotic amplitudes with $I^G(J^{PC})=1^-(1^{-+})$
belonging to the $\{10\}-\{\bar{10}\}$ representation of $SU(3)$ for the
inelastic reactions $\rho\pi\to\eta_8\pi$, $K^*\bar K\to\eta_8\pi$, and $\bar
K^*K\to\eta_8\pi$ as well as the amplitudes belonging to the octet
representation of $SU(3)$ for the reaction $\rho\pi\to\eta_0\pi$, $K^*\bar K
\to\eta_0\pi$, and $\bar K^*K\to\eta_0\pi$. In the next orders, these tree
amplitudes induce as well the $I^G(J^{PC})=1^-(1^{-+})$ exotic ones for the
elastic processes $\rho\pi\to\rho\pi$, $\eta\pi\to\eta\pi$, and so on. In
this connection it is of interest to consider the following $4\times4$ system
of scattering amplitudes for the coupled exotic channels of the reactions
$VP\to VP$, $VP\leftrightarrow PP$ and $PP\to PP$:\begin{eqnarray}T_{ij}=\left[
\begin{array}{llll}
T(\rho\pi\to\rho\pi) & T(\rho\pi\to\eta\pi) & T(\rho\pi\to\eta'\pi) &
T(\rho\pi\to K^*K)\\
T(\eta\pi\to\rho\pi) & T(\eta\pi\to\eta\pi) & T(\eta\pi\to\eta'\pi) &
T(\eta\pi\to K^*K)\\
T(\eta'\pi\to\rho\pi) & T(\eta'\pi\to\eta\pi) & T(\eta'\pi\to\eta'\pi) &
T(\eta'\pi\to K^*K)\\
T(K^*K\to\rho\pi) & T(K^*K\to\eta\pi) & T(K^*K\to\eta'\pi) & T(K^*K\to K^*K)\\
\end{array}\right]\,.\end{eqnarray} Here the subscripts $i,j=1,2,3,4$ are the
labels of the $\rho\pi$, $\eta\pi$, $\eta'\pi$, and $K^*K$ channels,
respectively, and the abbreviation $K^*K$ implies just the $\bar K^*K$ and $K^*
\bar K$ channels. The corresponding matrix of the $VPPP$ coupling constants
generated by the Lagrangian (3) has the form \begin{eqnarray}
h_{ij}=h\left[\begin{array}{cccc}
0 & \alpha & \beta & 0\\ \alpha & 0 & 0 & \gamma\\ \beta & 0 & 0 & \delta\\
0 & \gamma & \delta & 0\end{array}\right]\,,\end{eqnarray} where
\begin{eqnarray}
\alpha=\sqrt{\frac{1}{3}}\cos\theta_P-\frac{h'}{h}
\sqrt{\frac{2}{3}}\sin\theta_P\,,\ \ \
\beta =\sqrt{\frac{1}{3}}\sin\theta_P+\frac{h'}{h}
\sqrt{\frac{2}{3}}\cos\theta_P\,,\\
\gamma=\sqrt{\frac{2}{3}}\cos\theta_P+\frac{h'}{h}
\sqrt{\frac{1}{3}}\sin\theta_P\,,\ \ \
\delta=\sqrt{\frac{2}{3}}\sin\theta_P-\frac{h'}{h}
\sqrt{\frac{1}{3}}\cos\theta_P\,.\end{eqnarray}
In the following we shall consider three natural limiting cases: $(i)$ $h'=0
$, i.e., when all exotic amplitudes belong to the $\{10\}-\{\bar{10}\}$
representation of $SU(3)$, $(ii)$ $h=0$, i.e., when all exotic amplitudes
belong to the octet representation of $SU(3)$, and $(iii)$ $h'=h$, when the
original pointlike $VPPP$ interaction possesses nonet symmetry with respect to
the pseudoscalar mesons.

To satisfy the unitarity condition for the coupled channel amplitudes, we sum
up all the possible chains of the $s$-channel loop diagrams the typical examples
of which are shown in Fig. 1. Such an old-fashioned field theory way of the
unitarization is well known in the literature (see, for example, Refs.
[52-55,32]). Notice that in case $(i)$ and in case $(ii)$ the whole complex of
the unitarized amplitudes, in fact, can be constructed by using only the
amplitudes for the loop diagrams shown explicitly in
Fig. 1. The point is that in these cases the denominator of the corresponding
geometrical series for any channel turns out to be proportional to
the sum of diagrams $(b)$, $(c)$, $(d)$, and $(e)$ in Fig. 1, and the
loop diagrams of the type $(f)$ and $(g)$, or $(h)$ and $(i)$, play a role of
a ``priming"\ in the corresponding elastic channels like diagram $(a)$ in the
inelastic $\rho\pi\to\eta\pi$ channel. However, in case $(iii)$ the situation
is considerably more complicated.

Before summing the diagrams, let us make two remarks about the model itself.

First, generally speaking, the pointlike exotic contributions due to the $VP
PP$ interaction might be modified by the tree diagrams involving $V$ meson
exchanges if one takes into account the ``anomalous"\ Lagrangian for the $VVP$
interaction and the ordinary one for the $VPP$ interaction. However, such a
considerable complication of the original exotic amplitudes actually does not
lead to any new possibilities (or degrees of freedom) to obtain the
resonancelike behavior of the complete unitarized amplitudes. This only
burdens the model by additional technical difficulties and makes it much
less transparent in comparison with the one based only on the Lagrangian (3).
For example, after such a modification of the original exotic amplitudes, the
above-mentioned obvious unitarization scheme need be changed, say, by some
version of the Pad\'e approximation [27,28] because of the impossibility of the
direct calculation and summation of higher loops.

Second, the effective coupling constant $h$ occurring in the Lagrangian
(3) is not the unambiguously definite value in the theory with the ``anomalous"
\ chiral Lagrangians (comprehensive discussions of this point may be found
in Refs. [36-43]). Actually, one may only claim that it is not too
large in the scale defined by the combination $2g_{\rho\pi\pi}g_{\omega\rho\pi}
/m^2_\rho\approx284$ GeV$^{-3}$ [40,43]. Therefore, we consider the coupling
constants $h$ and $h'$ as free parameters of the model in the region of their
relatively small values.

The summing up of the loop diagram chains can be easily carried out by using
the matrix equation for the auxiliary amplitudes $\tilde T_{ij}$,
\begin{equation}\tilde T_{ij}=h_{ij}+h_{im}\Pi_{mn}\tilde T_{nj}\,,
\end{equation} which is shown schematically in Fig. 2. The auxiliary amplitudes
$\tilde T_{ij}$ pertain to the hypothetical case when all the particles in the
reactions are spinless, but otherwise they are the exact analogs of the
physical amplitudes $T_{ij}$ designated in Eq. (4). So in Eq. (8)
the matrix $h_{ij}$ is given by
Eqs. (5), (6), and (7), and $\Pi_{ij}$ is the diagonal matrix of the loops,
i.e., $\Pi_{ij}=\delta_{ij}\Pi_j$, where $\Pi_1$, $\Pi_2$, $\Pi_3$, and $\Pi
_4$ correspond to the four independent $s$-channel loops involving the $\rho\pi
$, $\eta\pi$, $\eta'\pi$, and $K^*K$ intermediate states, respectively (for a
moment, all for the spinless case). Notice that if all the particles are
spinless, then the $h_{ij}$ and $\Pi_{ij}$ in Eq. (8) are dimensionless, as 
well as the $\tilde T_{ij}$ themselves.
For the matrix elements $h_{im}\Pi_{mj}=h_{ij}\Pi_j$
it is convenient to introduce the following compact notation [look at Eq.
(5)]: \begin{eqnarray} h_{im}\Pi_{mj}=h_{ij}\Pi_j
=h\left[\begin{array}{cccc} 0 & \alpha_2 & \beta_3 & 0\\ \alpha_1 & 0 & 0 &
\gamma_4\\ \beta_1 & 0 & 0 & \delta_4\\ 0 & \gamma_2 & \delta_3 & 0\end{array}
\right]\,,\end{eqnarray} where $\alpha_1=\alpha\Pi_1$, $\beta_3=\beta\Pi_3$,
and so on. The solution of Eq. (8) has the form \begin{equation}\tilde
T_{ij}=[\,(\hat1-\hat h\hat\Pi)^{-1}\,]_{im}\,h_{mj}\,,\end{equation} where $
\hat1$ is the $4\times4$ identity matrix and the matrix $\hat h\hat\Pi$ is
defined by the relations of Eq. (9). Next, we define \begin{equation} \bar D=
\mbox{det}(\hat1-\hat h\hat\Pi)=1-h^2(\alpha_1\alpha_2+\beta_1\beta_3+\gamma_2
\gamma_4+\delta_3\delta_4)+h^4(\alpha_1\delta_4-\beta_1\gamma_4)(\alpha_2\delta
_3-\beta_3\gamma_2)\,.\end{equation} Let us now write down, as an example, the
explicit expressions for the amplitudes of the following five reactions:
\begin{equation}\tilde T(\rho\pi\to\rho\pi)=h^2[
\alpha\alpha_2+\beta\beta_3-h^2(\alpha\delta_4-\beta\gamma_4)(\alpha_2\delta_3-
\beta_3\gamma_2)]/\bar D\,,\end{equation} \begin{equation}\tilde T(\rho\pi\to
\eta\pi)=h\,[\alpha-h^2(\alpha\delta_3-\beta_3\gamma)\delta_4]/\bar D\,,
\end{equation}\begin{equation}\tilde T(\rho\pi\to\eta'\pi)=h\,[\beta+h^2(\alpha
_2\delta-\beta\gamma_2)\gamma_4]/\bar D\,,\end{equation}\begin{equation}\tilde
T(\rho\pi\to K^*K)=h^2[\alpha_2\gamma+\beta_3\delta]/\bar D\,.\end{equation}
\begin{equation}\tilde T(\eta\pi\to\eta\pi)=h^2[\alpha\alpha_1+\gamma
\gamma_4-h^2(\alpha\delta_3-\beta_3\gamma)(\alpha_1\delta_4-\beta_1\gamma_4)]
/\bar D\,,\end{equation} In cases $(i)$ and $(ii)$, the combination $h
^2(\alpha\delta-\beta\gamma)=0$ [see Eqs. (5), (6), and (7)], so that the
contributions proportional to that vanish in Eqs. $(10)-(16)$ and, as one can
see, all the formulas are essentially simplified [for example, the numerator in
Eq. (13) becomes simply equal to $h\alpha$, since, according to Eq. (9),
$h^2(\alpha\delta_3-\beta_3\gamma)=h^2(\alpha\delta-\beta\gamma)\Pi_3$].

Let us now take into account the spin of the particles. Consider the three
different processes
\begin{equation}\rho^0(k)+\pi^-(q_1)\to\rho^0(k')+\pi^-(q'_1)\,,\end{equation}
\begin{equation}\rho^0(k)+\pi^-(q_1)\to\eta(q_2)+\pi^-(q_3)\,,\end{equation}
\begin{equation}\eta(p)+\pi^-(q)\to\eta(q_2)+\pi^-(q_3).\end{equation}
Let $Q=k+q_1=k'+q'_1=q_2+q_3=p+q$ and $\,s=Q^2$. Straightforward
calculations with the help of the Lagrangian (3) of arbitrary terms of
the relevant diagram series results in the following Lorentz structures and
angular dependences for the corresponding physical amplitudes:
\begin{equation}T^{(\lambda',\lambda)}(\rho^0\pi^-\to\rho^0\pi^-)=\epsilon
_{\mu'\nu'\tau'\sigma}e_{(\lambda')}^{\mu'*}q'^{\nu'}_1k'^{\tau'}\,\epsilon
^\sigma_{\,\,\mu\nu\tau}e_{(\lambda)}
^\mu q_1^\nu k^\tau\,\,\tilde T'(\rho\pi\to\rho\pi)=\end{equation}$$
(\delta_{\lambda,+1}+\delta_{\lambda,-1})(\delta_{\lambda',+1}+\delta_{\lambda'
,-1})(s|\vec q_1|^2/2)(\lambda\lambda'+\cos\theta)\,\tilde T'(\rho\pi\to\rho\pi
)\,,$$\begin{equation}T^{(\lambda)}(\rho^0\pi^-\to\eta\pi^-)=\epsilon
_{\mu\nu\tau\sigma}e_{(\lambda)}^\mu q_1^\nu q_2^\tau q_3^\sigma\,\,\tilde T'(
\rho\pi\to\eta\pi)=\end{equation}$$-(\delta_{\lambda,+1}+\delta_{\lambda,-1})
i\sqrt{s/2}|\vec q_1||\vec q_3|\sin\theta\,\tilde T'(\rho\pi\to\eta\pi)\,,
$$\begin{equation}T(\eta\pi^-\to\eta\pi^-)=|\vec q|^2\cos\theta\,\tilde T'(
\eta\pi\to\eta\pi)\,,\end{equation}where $\lambda$ ($\lambda'$) is the initial
(final) $\rho$ meson helicity and $\theta$ is the angle between the momenta of
the initial and final pions in the reaction c.m. system. Certainly the
dimensions of all physical amplitudes $T$ in Eqs. $(20)-(22)$ are the same:
the amplitudes are dimensionless. At the same time, as is seen from Eqs. $(20)-
(22)$, the invariant amplitudes $\tilde T'$ have different dimensions in
the $VP\to VP$, $VP\to PP$, and $PP\to PP$ channels. These invariant
amplitudes are obtained directly from the corresponding auxiliary amplitudes $
\tilde T$ [see Eqs. $(10)-(16)$] by substituting the physical dimensional
coupling constants $h$ and $h'$ from the Lagrangian (3) and the following
expressions for the $p$-wave loop integrals:
\begin{eqnarray}\Pi_{i}=\frac{1}{16\pi}\,\frac{2}{3}\,F
_i\times\left\{\begin{array}{r}4s,\  i=1,4\ \, (VP\ \mbox{loops})\,,\\ 1,\  i=2
,3\ \, (PP\ \mbox{loops})\,,\\ \end{array}\right.\end{eqnarray} where
\begin{equation}F_i=C_{1i}+sC_{2i}+\frac{s^2}{\pi}\int\limits_{m^2_{i+}}^
{\infty}\frac{[P_i(s')]^3\ ds'}{\sqrt{s'}\,s'^{\,2}(s'-s-i\varepsilon)}=
C_{1i}+sC_{2i}\,+\end{equation} $$\frac{(s-m_{i+}^2)^{3/2}(s-m_{i-}^2)^{3/2}}
{8\pi s^2}\,\left[\,\ln\,\left(\frac{\sqrt{s-m_{i-}^2}-\sqrt{s-m_{i+}^2}}{
\sqrt{s-m_{i-}^2}+\sqrt{s-m_{i+}^2}}\,\right)+i\pi\,\right]+$$
$$\frac{1}{4\pi}\left\{\frac{1}{2m_{i+}m_{i-}}\,\ln\left(\frac{m_{i+}-m_{i-}}{m
_{i+}+m_{i-}}\right)\,\left[\frac{m_{i+}^4m_{i-}^4}{s^2}-\frac{3m_{i+}^2m_{i-}
^2}{2s}(m_{i+}^2+m_{i-}^2)\,+\right.\right.$$
$$\left.\frac{3}{8}(m_{i+}^4+m_{i-}^4+
6m_{i+}^2m_{i-}^2)+\frac{s(m_{i+}^2+m_{i-}^2)}{16m_{i+}^2m_{i-}^2}(m_{i+}
^4-10m_{i+}^2m_{i-}^2+m_{i-}^4)\right]+$$
$$\left.\frac{m_{i+}^2m_{i-}^2}{2s}-\frac{5}{8}(m_{i+}^2+m_{i-}^2)+\frac{s(3m
_{i+}^4+3m_{i-}^4+38m_{i+}^2m_{i-}^2)}{48m_{i+}^2m_{i-}^2}\right\}\,.$$ Here $P
_i(s)=[(s-m_{i+}^2)(s-m_{i-}^2)/(4s)]^{1/2}\,$, $m_{i+}$ ($m_{i-}$) is the sum
(the difference) of the particle masses in channel $i$, and $C_{1i}$ and $C_{2i
}$ are the subtraction constants. Note that the expression (24) is valid for $s
\geq m^2_{i+}$. In the regions $m^2_{i-}<s<m^2_{i+}$ and $s\leq m^2_{i-}$, it
changes according to analytic continuation [56].

\begin{center}{\bf IV. ANALYSIS OF THE POSSIBLE RESONANCE PHENOMENA}
\end{center}

First of all let us note that a number of free parameters in the present model
can be reduced essentially, leaving its potentialities almost unchanged. So
we shall assume that $C_{11}=C_{14}$, $C_{21}=C_{24}$ for the $VP$ loops and $C
_{12}=C_{13}$, $C_{22}=C_{23}$ for $PP$ loops. Moreover, near a feasible
resonance, the smooth $s$ dependence of the combinations $C_{1i}+sC_{2i}$ is
not of crucial importance. Thus, as the essential free parameters we can leave
only the $C_{11}$ and $C_{12}$ ones, setting $C_{21}=C_{22}=0$. Just this will
be done in most variants considered below. A simplest way to discover ``by
hand"\ a possible
resonance situation is that to find zero of Re$(\bar D)$ at fixed values of $h
$, $h'$, and $\sqrt{s}$, for example, at $\sqrt{s}=1.43$ GeV [see Eqs. $(11)-
(16)$]. In so doing the left free subtraction constants $C_{11}$ and $C_{12}$
are not uniquely determined. For example, in cases $(i)$ and $(ii)$, the
condition Re$(\bar D)=0$ gives only a relation of the type $C_{12}=(\xi_1+\xi
_2C_{11})/(\xi_3+\xi_4C_{11})$, where $\xi_i$ are the known numbers. However,
this is not the weak point of the model; on the contrary, this allows the
shapes of the resonance curves and the relations between the absolute cross
section values in the different channels to be easily changed by changing $C
_{11}$.

According the detailed analysis performed in Refs. [36-40,43], the acceptable
tentative values of the parameter $\tilde h\equiv F^3_\pi h$ (where $F_\pi
\approx130$ MeV) lie within the range $|\tilde h|\leq0.4$. To illustrate the
existence of the resonance phenomena in our toy model we are guided by the
values of $\tilde h$ (and $\tilde h'\equiv F^3_\pi h'$) near 0.1. We would like
particularly to emphasize that, in fact, the resonance phenomena are possible
in the present model for any $|\tilde h|\leq0.4$. However, as $|\tilde h|$
(and/or $|\tilde h'|$) increases from 0.1 to 0.4, the distinct resonancelike
enhancements in the reaction cross sections move into the region $\sqrt{s}
\approx1-1.3$ GeV. Note that the unitarized amplitudes essentially depend on
the second and fourth powers of coupling constants and therefore are very
sensitive to changes of $|\tilde h|$ and $|\tilde h'|$.

In Figs. 3 and 4, we show the typical energy dependences, which occur in our
model for cases $(i)$, $(ii)$, and $(iii)$, for the four reaction cross
sections $\sigma(\rho^0\pi^-\to\rho^0\pi^-)$, $\sigma(\rho^0\pi^-\to\eta
\pi^-)$, $\sigma(\rho^0\pi^-\to\eta'\pi^-)$, and $\sigma(\rho^0\pi^-\to K^{*0}K
^-)$ and for the phases of the $\rho\pi\to\rho\pi$ and $\rho\pi\to\eta\pi$
amplitudes (note that the inelastic amplitude $\rho\pi\to\eta\pi$ is defined
only up to the sign). 
Figure 3, together with Table I, and Fig. 4, together with Table II,
illustrate the resonance effects when they concentrate mainly in the regions $
\sqrt{s}\approx1.3-1.4$ GeV and $\sqrt{s}\approx1.5-1.6$ GeV, respectively. As
a rule, the channel $\rho\pi\to\eta\pi$ is dominant in case $(i)$, when all
considered amplitudes belong to the $\{10\}-\{\bar{10}\}$ representation of $SU
(3)$. In case $(ii)$, when the amplitudes belong to the octet representation of $SU(
3)$, the main
channels are the $\rho\pi\to\rho\pi$ and $\rho\pi\to\eta'\pi$ ones. In case $(i
ii)$, when $h'=h$ and the Lagrangian (3) possesses additional nonet symmetry,
the cross sections for all channels except the $K^*K$ one turn out to be
comparable, and the general situation is rather complicated. 
The branching ratios of the presented resonancelike enhancements to the 
$\rho^0\pi^-\to(\rho^0\pi^-$, $\eta\pi^-$, $\eta'\pi^-$, $K^{*0}K^-)$ channels
are listed in Tables I and II. Such characteristics for the complicated broad
resonance structure can be defined as follows. For example, $B(\rho^0\pi^-)=
\bar\sigma(\rho^0\pi^-\to\rho^0\pi^-)/\Sigma$, $B(\eta\pi^-)=\bar\sigma(\rho^0
\pi^-\to\eta\pi^-)/\Sigma$, etc., where $\Sigma=2\bar\sigma(\rho^0\pi^-\to\rho
^0\pi^-)$ + $\bar\sigma(\rho^0\pi^-\to\eta\pi^-)$ + $\bar\sigma(\rho^0\pi^-\to
\eta'\pi^-)$ + $2\bar\sigma(\rho^0\pi^-\to K^{*0}K^-)$ and every $\bar\sigma$ 
is the integral of the corresponding cross section over the $\sqrt{s}$ range 
from 1.2 to 1.8 GeV, where an enhancement concentrates. 

Let us now compare the cross section values shown in Figs. 3 and 4 with those
of $a_2(1320)$ resonance production. Using the tabular branching ratios [1] we
get $\sigma(\rho^0\pi^-\to a_2\to\rho^0\pi^-)\approx5.7$ mb and $\sigma(\rho^0
\pi^-\to a_2\to
\eta\pi^-)\approx2.36$ mb at $\sqrt{s}=m_{a_2}=1.32$ GeV. Taking also
into account the ratio of the factors $(2J+1)/|\vec k|^2$ for the $a_2(1320)$
resonance and for the $J=1$ enhancement found at $\sqrt{s}\approx1.3-1.4$ GeV,
or at $\sqrt{s}\approx1.5-1.6$ GeV, we can conclude that we are certainly
dealing with the resonancelike behavior of the $I^G(J^{PC})=1^-(1^{-+})$
exotic waves in the region $1.3\leq\sqrt{s}\leq1.6$ GeV, at least, in the $\rho
\pi$, $\eta\pi$, and $\eta'\pi$ channels.

The most appreciable manifestation of an $I^G(J^{PC})=1^-(1^{-+})$ exotic state
in the mass region $1.3-1.4$ GeV has been observed in the $\eta\pi^0$ channel
in the charge exchange reaction $\pi^-p\to\eta\pi^0n$ at 32, 38, and 100 GeV/$c
$ [11] (currently the exotic states of such a type are denoted most commonly as
$\pi_1$). It was found that the intensity of the $\pi_1$ signal at its
maximum in this reaction is only 3.5 times smaller than the corresponding
intensity of the $a_2(1320)$ signal. It is very essential that $a_2(1320)$
production and $\pi_1$ production both proceed in this case via a single
mechanism, namely, via the Reggeized $\rho$ exchange. If the $\pi_1$ really
represents a complicated structure of the $qq\bar q\bar q$ or $q\bar qg$ type,
then the fact that the $\pi_1$ production cross section in the charge exchange
reaction has been found fully comparable with that of the conventional
$q\bar q$ resonance $a_2(1320)$ is certainly strong evidence for
the resonance nature of the observed exotic signal.

On the other hand, the intensity of $\pi_1$ production in the $\eta\pi^-$
channel in the reactions $\pi^-p\to\eta\pi^-p$ at 37 GeV/$c$ [3] and 18 GeV/$c$
[12] has been found to be about 15 and, respectively, 30 times smaller than
the $a_2(1320)$ production intensity.
This is evidently due to the more complicated mechanism of the reaction $\pi^-p
\to\eta\pi^-p$ than that of the charge exchange reaction. In fact, there are
three competing Regge exchanges with natural parity in this reaction: the
$\rho$ exchange, the $f_2$ exchange, and the Pomeron one.
Also, as is known, the last two are dominant in
the case of $a_2(1320)$ production [57]. Note that $\pi_1$ production can
proceed via the Pomeron mechanism only owing to the octet component of the $\pi
_1$. However, if this component is small, that is, if the $\pi_1$ belongs
mainly to the $\{10\}-\{\bar{10}\}$ representation of $SU(3)$, then $\pi_1$
production via Pomeron exchange has to be suppressed.

Another opportunity to observed $\pi_1$ and $a_2(1320)$ formation with
comparable cross sections appears by using photoproduction (and
electroproduction) processes, for example, $\gamma p\to\rho^0\pi^-\Delta^{++}
\to\pi^+\pi^-\pi
^-\Delta^{++}$, $\gamma p\to\rho^0\pi^+n\to\pi^+\pi^-\pi^+n$, $\gamma p\to\eta
\pi^+n$, and so on, which go at low momentum transfer mainly via the Reggeized
one-pion exchange mechanism. Indeed, the existing data on the reactions $\gamma
p\to\rho^0\pi^-\Delta^{++}\to\pi^+\pi^-\pi^-\Delta^{++}$ and $\gamma p\to\rho^0
\pi^+n\to\pi^+\pi^-\pi^+n$ show a clear signature of the $a_2(1320)$ resonance
and the appreciable enhancements in the $3\pi$ mass spectra in the range $1.5-2
$ GeV [58]. However, they do not yet allow certain conclusions to be made
concerning the presence of the exotic wave in the $\rho\pi$ system and further
investigations of the above reactions are needed.

At the present time an extensive program of the search for the exotic $\pi_1$
states in photoproduction experiments with high statistics and precision is
planned for the Jefferson Laboratory [6,14,15,18-20]. A careful
study of the $\pi_1\to\gamma\pi$ radiative decays in hadroproduction from
nuclei via the Primakoff one-photon exchange mechanism is also planned with the
CERN COMPASS spectrometer [59]. A collection of the data on $\pi_1$
photoproduction, electroproduction, and hadroproduction and on the decays of 
the $\pi_1$ into $\rho\pi$ will also allow for the first time to verify the 
vector meson dominance model for states with exotic quantum numbers.

Summarizing we conclude that our calculation gives a further new reason in
favor of the plausibility of the existence of an explicitly exotic resonance
with $I^G(J^{PC})=1^-(1^{-+})$ in the mass range $1.3-1.6$ GeV.
Currently two exotic states at 1.4 and 1.6 GeV are extensively discussed in
the literature [7-20]. In our scheme one does not succeed in simultaneously
generating both the 1.4 and the 1.6 resonances, although the variants with a 
``fine structure"\ exist [see, for example, the cross sections for case $(iii)$
in Figs. 3 and 4]. Such a ``fine structure"\ will be smoothed by the 
experimental resolution and we cannot certainly say about two resonances. The 
question may be raised as to whether this is a crucial result. It is not 
improbable that the inclusion of the $b_1\pi$ and $f_1\pi$ channels, where the
exotic signals have also been found, can change the situation. However, the 
issue of the additional $b_1\pi$ and $f_1\pi$ channels remains open in the 
effective chiral Lagrangian approach. Notice also that at present the situation
with the two exotic resonances at 1.4 and 1.6 GeV is not yet finally arranged. 

\begin{center}{\bf ACKNOWLEDGMENTS}\end{center}
The present work was supported in part by grant INTAS-RFBR IR-97-232.

\vspace*{2cm}
{\bf TABLE I.} The parameter values of the model for the curves in
Fig. 3. $C_{11}$ and $C_{12}$ are in GeV$^2$; the other parameters are
dimensionless; $C_{21}=C_{22}=0$ in all cases. Also, in the four right columns,
the branching ratios of the resonancelike enhancement obtained to the partial 
channels are presented.
\begin{center}
\begin{tabular}{|c|c|c|c|c|c|c|c|c|} \hline Cases & $F^3_\pi h$ & $F^3_\pi 
h'$ & $C_{11}$  & $C_{12}$  & $B(\rho^0\pi^-)$ & $B(\eta\pi
^-)$ & $B(\eta'\pi^-)$ & $B(K^{*0}K^-)$ \\[0.1cm] \hline
$(i)$      & 0.10746 & 0       & 0.17      & 1.25      &
0.2377 & 0.3614 & 0.0125 & 0.0754 \\
$(ii)$     & 0       & 0.10746 & 0.34      & 0.67      &
0.3259 & 0.0890 & 0.1924 & 0.0289 \\
$(iii)$    & 0.10746 & 0.10746 & 0.49      & 0.50      &
0.2534 & 0.3619 & 0.1276 & 0.0019 \\ \hline
\end{tabular}\end{center}
\vspace*{2cm}
{\bf TABLE II.} The parameter values of the model for the curves in
Fig. 4. $C_{11}$ and $C_{12}$ are in GeV$^2$; the other parameters are
dimensionless; $C_{21}=C_{22}=0$ in cases $(i)$ and $(ii)$ and 
$C_{21}=C_{22}=0.11$ in case $(iii)$. 
Also, in the four right columns, the branching ratios of the
resonancelike enhancement obtained to the partial channels are presented.
\begin{center}
\begin{tabular}{|c|c|c|c|c|c|c|c|c|} \hline Cases & $F^3_\pi h$ & $F^3_\pi 
h'$ & $C_{11}$  & $C_{12}$  & $B(\rho^0\pi^-)$ & $B(\eta\pi
^-)$ & $B(\eta'\pi^-)$ & $B(K^{*0}K^-)$ \\[0.1cm] \hline
$(i)$      & 0.10746 & 0       & 0.18      & 0.76      &
0.1616 & 0.4634 & 0.0217 & 0.0959 \\
$(ii)$     & 0       & 0.08417 & 0.33      & 0.78      &
0.3032 & 0.0804 & 0.2184 & 0.0474 \\
$(iii)$    & 0.10746 & 0.10746 & 0.11      & 0.11      &
0.2429 & 0.3686 & 0.1356 & 0.0050\\ \hline
\end{tabular}\end{center}

\newpage\begin{figure}\centerline{\epsfysize=8.5in\epsfbox{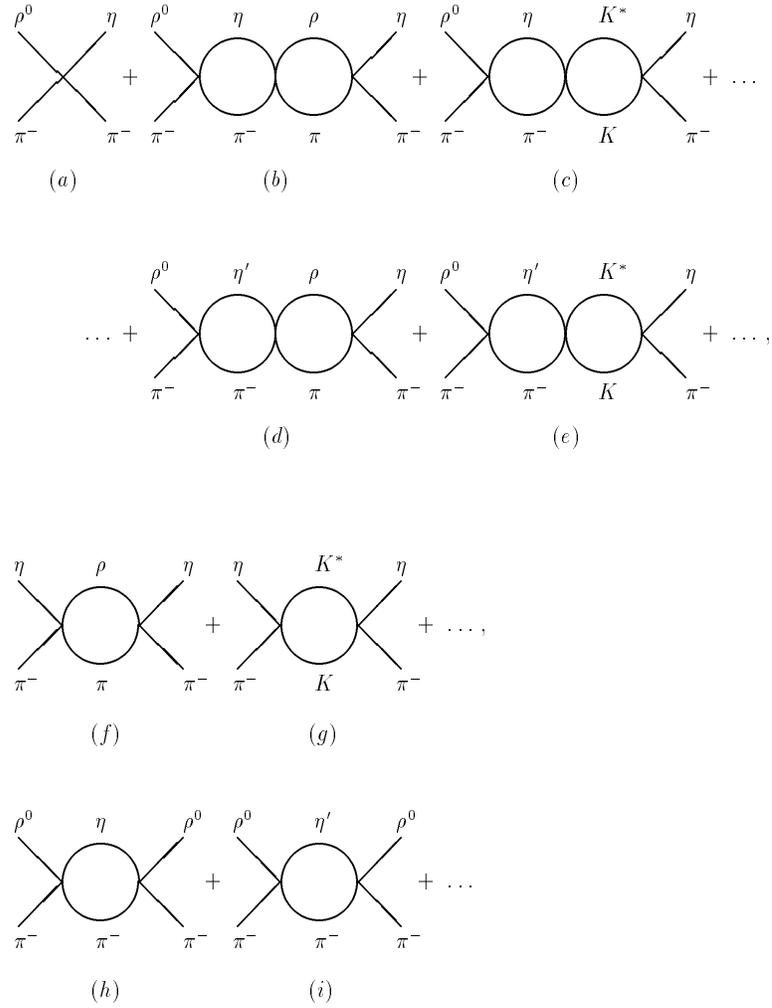}}\caption{
\  The examples of the diagrams which are summed to obtain the
unitarized amplitudes in coupled channels.}\end{figure}

\newpage\begin{figure}\centerline{\epsfysize=8.5in\epsfbox{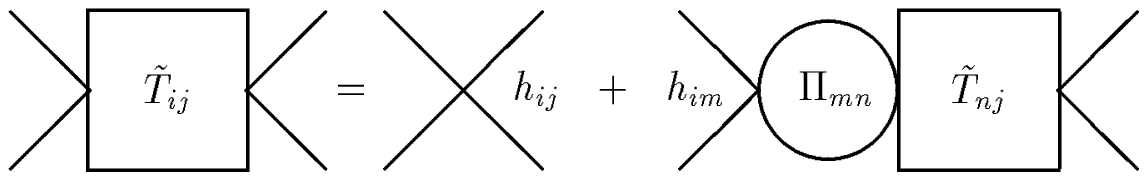}}\caption{
\  The auxiliary equation for the summing up of the diagram
series some examples of which are shown in Fig. 1.}\end{figure}

\newpage\begin{figure}\centerline{\epsfysize=8.3in\epsfbox{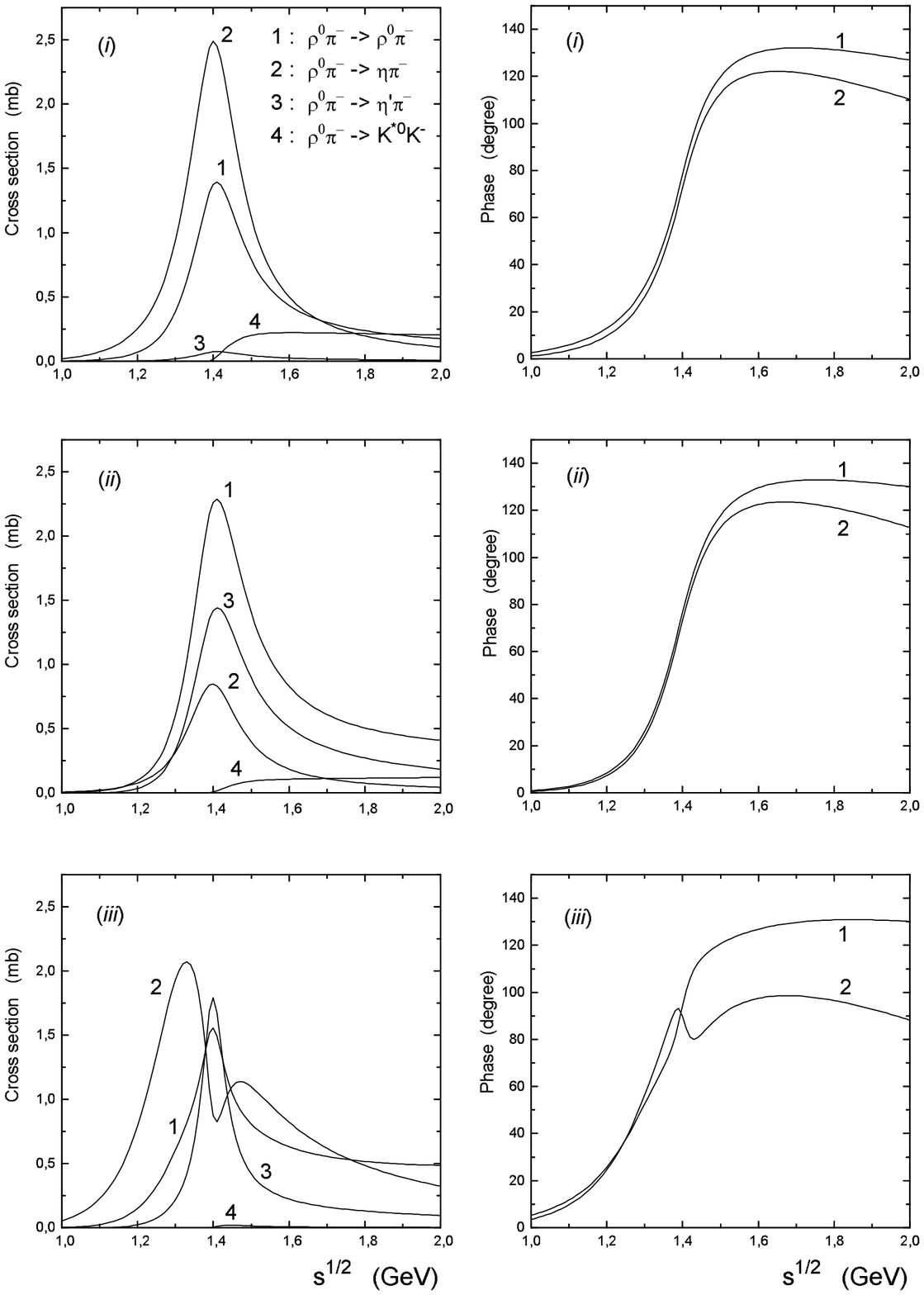}}\caption{
\  The cross sections of the reactions $\rho^0\pi^-\to\rho^0\pi^-$, $\rho^0\pi
^-\to\eta\pi^-$, $\rho^0\pi^-\to\eta'\pi^-$, and $\rho^0\pi^-\to K^{*0}K^-$ and
the phases of the $\rho\pi\to\rho\pi$ and $\rho\pi\to\eta\pi$ amplitudes
for cases $(i)$, $(ii)$, and $(iii)$. The correspondence between the
curve numbers and the reaction channels is shown just in the figure. The 
values used of the parameters are listed in Table I.}\end{figure}

\newpage\begin{figure}\centerline{\epsfysize=8.3in\epsfbox{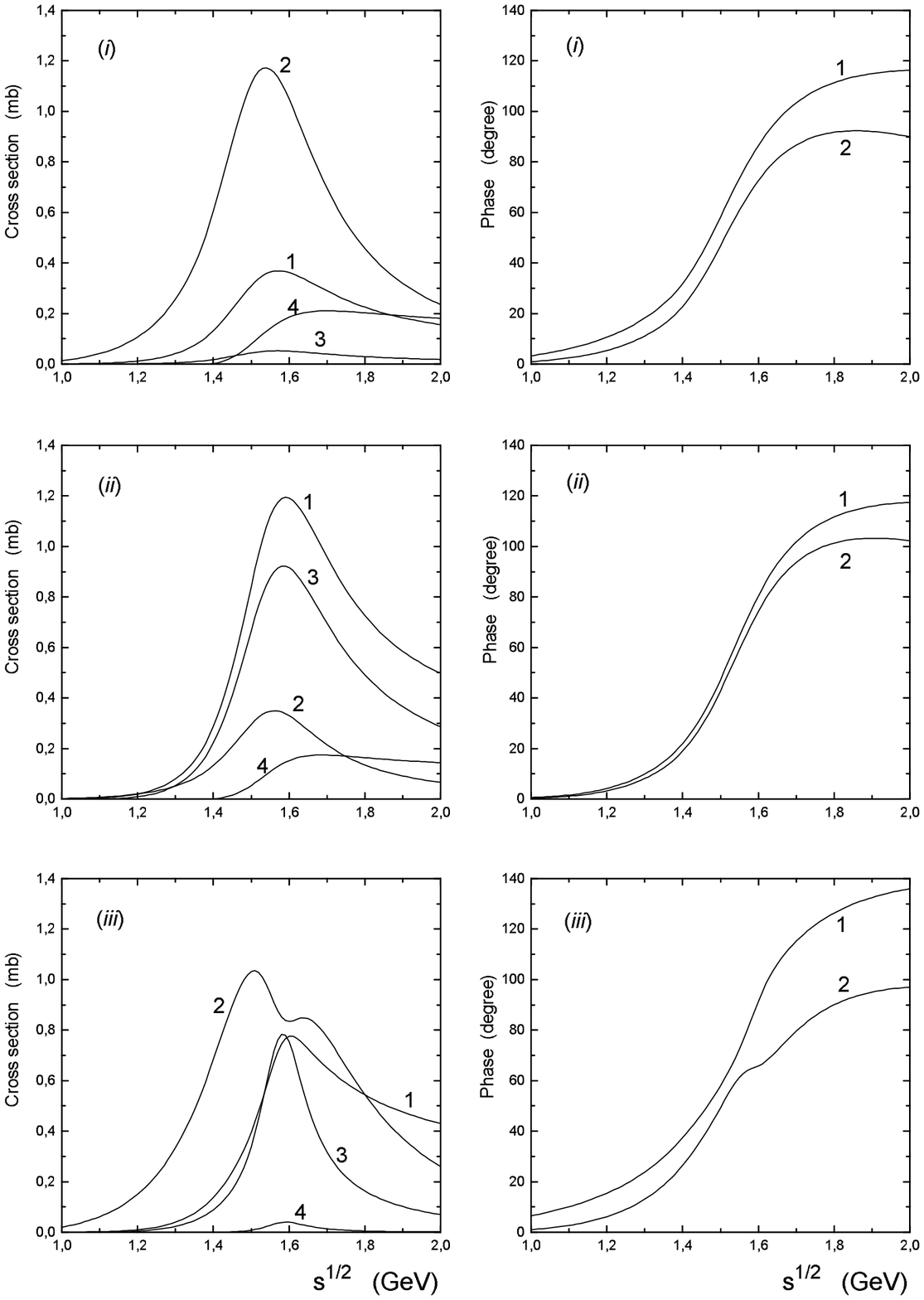}}\caption{
\  The cross sections of the reactions $\rho^0\pi^-\to\rho^0\pi^-$, $\rho^0\pi
^-\to\eta\pi^-$, $\rho^0\pi^-\to\eta'\pi^-$, and $\rho^0\pi^-\to K^{*0}K^-$ and
the phases of the $\rho\pi\to\rho\pi$ and $\rho\pi\to\eta\pi$ amplitudes
for cases $(i)$, $(ii)$, and $(iii)$. The correspondence between the
curve numbers and the reaction channels is the same as in Fig. 3. The values 
used of the parameters are listed in Table II.}\end{figure}\end{document}